\documentclass[a4paper,fleqn]{article}

\usepackage[numbers]{natbib}
\usepackage{amsmath}
\usepackage{graphicx}
\usepackage{enumerate}
\usepackage{url} % not crucial - just used below for the URL 
\usepackage{bm}
\usepackage{amssymb} % has symbol for reals
\usepackage{mathtools} % has a big cartesian product symbol
\usepackage{setspace}
\usepackage{algorithm}
\usepackage{algorithmic}
\usepackage{xspace}
\usepackage{multirow}
\DeclareMathOperator*{\argmin}{argmin}
\DeclareMathOperator*{\argmax}{argmax}

\newcommand{\eg}{e.g.\@ }

\begin{document}

\title {Revisiting the effect of greediness on the efficacy of exchange algorithms for generating exact optimal experimental designs}  

%\tnotemark[] 

%\tnotetext[<tnote number>]{} 


\author{William T. Gullion and Stephen J. Walsh \\ Utah State University}

\maketitle
\date

\begin{abstract}
Coordinate exchange (CEXCH) is a popular algorithm for generating exact optimal experimental designs. The authors of CEXCH advocated for a highly greedy implementation - one that exchanges and optimizes single element coordinates of the design matrix. We revisit the effect of greediness on CEXCHs efficacy for generating highly efficient designs. We implement the single-element CEXCH (most greedy), a design-row (medium greedy) optimization exchange, and particle swarm optimization (PSO; least greedy) on 21 exact response surface design scenarios, under the $D$- and $I-$criterion, which have well-known optimal designs that have been reproduced by several researchers. We found essentially no difference in performance of the most greedy CEXCH and the medium greedy CEXCH. PSO did exhibit better efficacy for generating $D$-optimal designs, and for most $I$-optimal designs than CEXCH, but not to a strong degree under our parametrization. This work suggests that further investigation of the greediness dimension and its effect on CEXCH efficacy on a wider suite of models and criterion is warranted. 
\end{abstract}


% Keywords
\\

\newpage

% Main text
\section{Introduction}\label{}
\cite{cexch} published the coordinate exchange (CEXCH) algorithm which enabled efficient generation of \textit{exact} optimal experimental designs (DoE) with fully continuous factor levels. Based on previous studies, the authors suggested that algorithm performance increased as the algorithm's degree of greediness increased. Greedy in this sense means exchanging individual scalar elements of a design matrix, seeking a local optimizer for each coordinate in sequence, as opposed to optimizing multiple design coordinates, full design points, or all elements of the design in an effort to find the global optimizer. In this paper we benchmark two versions of exchange algorithms each representing 2 different levels of greediness, on 21 well-known exact optimal design scenarios under two optimality criterion. We compare the results to previously published work which used particle swarm optimization (PSO) to generate optimal designs \citep{walsh, walsh1}. PSO is a fully non-greedy optimizer that has performed very well in generating optimal designs for various scenarios over the last decade and is rapidly gaining in popularity for solving optimal DoE problems.

\section{Survey of Algorithms for Generating Exact Optimal Designs}\label{}
The following sections contain a brief literature survey of two main types of algorithms researchers appeal to for generating optimal DoEs: exchange algorithms and meta-heuristics. 

\subsection{Exchange Algorithms}
The CEXCH of \cite{cexch} appears to be the most popular optimal design algorithm for both application and academic research \citep{odt, cexch, gIopt, rodman, saleh, stallrich}. This popularity, in part, appears to be due to CEXCHs simplicity and ease of implementation. CEXCH proceeds by generating a completely random starting design.  Each design coordinate is exchanged individually via some mechanism and the exchange that improves the optimality criterion the most is retained. The result is a \textit{locally} (relative to the random starting point) optimal design. To increase the chances of finding either a highly efficient design or the globally optimal design, several authors recommend applying CEXCH several thousand times to each design problem \cite{odt}.

\subsection{Particle Swarm Optimization}
In contrast to CEXCH, PSO attempts to search the space of candidate designs more globally \citep{walsh, walsh1, walsh2}. PSO tends to cost more than CEXCH but is highly robust to entrapment in local optima and therefore can have a higher probability of find the globally optimal design. PSO is a relative newcomer to optimal design. \cite{chen2} discuss PSO-generated latin-hypercube designs. \cite{chen1, joseph} use PSO to generate space-filling designs. \cite{lukemire, chen3, chen4} apply PSO to find optimal designs for non-linear models. PSO-generation of optimal designs for mixture experiments is provided in \cite{wong}.
\cite{psorev} provide a comprehensive review of PSO for generating continuous optimal designs. The efficacy of PSO for generating exact optimal designs was demonstrated in \cite{walsh1}. New $G$-optimal designs were efficiently generated for small-exact response surface scenarios by an adaption of PSO with the local-communication topology \citep{walsh2}. \cite{walsh3} used PSO to great effect to study the trade-offs of the $G$- and $I$-criterion (prediction variance designs) via a multi-objective optimization. A comprehensive review of PSO for solving the exact optimal design problem is provided in \cite{walsh}.

\section{Optimal Design for Small-Exact Response Surface Scenarios}
We denote an exact $N$-run (rows) experimental design as $\mathbf{X} = \begin{pmatrix} \mathbf{x}_1 & \mathbf{x}_2 & \hdots & \mathbf{x}_N \end{pmatrix}'$. Each row $\mathbf{x}'_i$ contains the settings of $K$ (columns) experimental factors. As in RSM, we consider the standardized design space, the hypercube, denoted  $\mathbf{x}'_i \in \mathcal{X} = [-1, 1]^K$ \citep{rsm}. We consider the second-order linear model which has $p = {K + 2 \choose 2}$ linear coefficient parameters. In scalar form, the second-order model is written $y = \beta_0 + \sum_{i = 1}^K\beta_ix_i + \sum_{i=1}^{K-1}\sum_{j = i + 1}^K\beta_{ij}x_i x_j + \sum_{i=1}^K \beta_{ii}x_i^2 + \epsilon$. Let the $N\times p$ matrix $\mathbf{F}$ represent the model matrix with rows given by the expansion vector $\mathbf{f}'(\mathbf{x}'_i) = (1 \quad x_{i1} \quad \hdots \quad x_{i2} \quad x_{i1}x_{i2} \quad \hdots \quad x_{i(K-1)}x_{iK} \quad x_{i1}^2 \quad \hdots \quad x_{iK}^2)$. Note that $\mathbf{F}$ is an explicit function of design $\mathbf{X}$. Thus, the linear model is $\mathbf{y} = \mathbf{F}\bm{\beta} + \bm{\epsilon}$ where we assume $\bm{\epsilon}\sim \mathcal{N}_N(\mathbf{0}, \sigma^2 \mathbf{I}_N)$. The ordinary least squares estimator of $\bm{\beta}$ is $\hat{\bm{\beta}} = (\mathbf{F}'\mathbf{F})^{-1}\mathbf{F}'\mathbf{y}$ which has variance $\text{Var}(\hat{\bm{\beta}}) = \sigma^2(\mathbf{F}'\mathbf{F})^{-1}$. The total information matrix for $\bm{\beta}$, $\mathbf{M}(\mathbf{X}) = \mathbf{F}'\mathbf{F}$, plays an important role in optimal DoE---all optimal design objective functions are functions of this matrix. The practitioner must choose a design from $\mathcal{X}^N$ to implement. An optimality criterion $f$ is used score candidate design quality. An optimization algorithm is required to search $\mathcal{X}^N$ under $f$ for high-quality designs. In this paper we consider the $D$- and $I$-criterion described in the next sections.

\subsection{$D$-optimal Designs}
The $D$-score of an arbitrary design $\mathbf{X}$ is the deteriminant of the inverse of the information matrix $D(\mathbf{X}) = |(\mathbf{F}'\mathbf{F})^{-1}|$. Thus the $D$-optimal design is	$\mathbf{X}^*_D  := \argmin_{\mathbf{X} \in \mathcal{X}^N} D(\mathbf{X})$. $D$-optimal designs are popular choices for screening experiments, and appear to be the most widely studied optimal DoE criterion \citep{cexch, cexch2, odt, adt, stallrich}.

\subsection{$I$-optimal Designs}
$I$-optimal designs are those that minimize the average prediction variance over $\mathcal{X}$. The scaled prediction variance is often defined as $\text{SPV}(\mathbf{x}'|\mathbf{X})  = N \mathbf{f}'(\mathbf{x}')(\mathbf{F}'\mathbf{F})^{-1}\mathbf{f}(\mathbf{x}')$ \citep{rodman, jobo1}. The $I$-criterion for candidate design $\mathbf{X}$ is defined as $I(\mathbf{X}) = \frac{1}{V} \displaystyle\int_{\mathcal{X}} \text{SPV}(\mathbf{x}'|\mathbf{X}) \mathbf{dx}' = \frac{N}{V} \text{tr}\biggl\{(\mathbf{F}'\mathbf{F})^{-1} \displaystyle\int_{\mathcal{X}} \mathbf{f}(\mathbf{x}')\mathbf{f}'(\mathbf{x}')\mathbf{dx}'\biggr\}$ where $V = \displaystyle\int_{\mathcal{X}} \mathbf{dx}'$ yields the volume of the design space $\mathcal{X}$. The $I$-optimal design is defined as: $\mathbf{X}^*_I :=  \argmin_{\mathbf{X}\in \mathcal{X}^N} I(\mathbf{X})$. $I$-optimal designs are popular RSM designs due to their focus on giving predictions with minimized variance. 

\subsection{Second-Order Model: Best-Known Exact Optimal Designs}
\cite{jobo1} applied a Genetic Algorithm (GA) to generate candidate exact optimal designs for 21 distinct design scenarios (all second-order model) for $K = 1$ ($N = 3, 4, 5, 6, 7, 8, 9$), $K = 2$ ($N = 6, 7, 8, 9, 10, 11, 12$), and $K = 3$ ($N = 10, 11, 12, 13, 14, 15, 16$) design factors and four optimality criteria: the $D$-, $A$-, $I$-, and $G$-criterion. These designs have been reproduced by several authors, see \eg \cite{rodman, saleh, gIopt, walsh, walsh1, walsh2, walsh3}. Thus, the catalog proposed by \cite{jobo1} has become a domain-standard dataset of exact optimal designs that is highly used for algorithm benchmarking. We adopt these designs as ground-truth for the purpose of this study.

\section{Implemented Algorithms}\label{}
For each of the 21 design scenarios in \cite{jobo1}, we ran an element-wise (fully greedy) and a row-wise (less greedy) EXCH algorithm $n_{\text{run}} = 10,000$ independent times. For PSO, the most non-greedy algorithm in this comparative study, we take advantage of a previously published benchmarking database for these specific design scenarios presented and described in \cite{walsh, walsh1} who published performance data on $n_{\text{run}} = 140$ independent PSO runs. These authors demonstrated high efficacy of PSO even with this number of runs. Brief descriptions of the CEXCH algorithms follow. 

\subsection{Element-wise CEXCH}
This is the most greedy algorithm in this study. Early versions of CEXCH used the \textit{addition}, \textit{deletion}, and \textit{delta} update functions which identify candidate coordinates to add/delete and measure the effect of these changes on the candidate optimal design \citep{cexch2, cexch, stallrich}. These update functions, however, are only mathematically tractable for simpler design problems and optimality criteria, such as the $D$- and $A$-criterion \citep{cexch, stallrich}. Therefore, it is common practice to implement CEXCH by adapting a univariate optimization algorithm, as shown in line \ref{optim:kvar:calc} of Algorithm \ref{alg:cexch}, as the proposal mechanism. For example, authors \cite{rodman, odt} apply Brent's minimization algorithm as the mechanism for proposing coordinate exchanges. In our code, we used the \texttt{BOBYQA} optimization algorithm (univariate) in \texttt{Julia}'s \texttt{NLOpt} package \citep{julia, nlopt}. In line \ref{optim:kvar:calc}, $x_{ij}$ represents the $ij$th element of candidate design $\mathbf{X}$, while $\mathbf{X}_{-ij}$ represents the candidate design \textit{without} the $ij$th element. Thus the univariate optimization attempts to find the value $x_{ij}$ that improves the criterion conditioned on all other design coordinates $\mathbf{X}_{-ij}$ remaining fixed. The procedure proceeds by looping through the rows and columns of $\mathbf{X}$ until a full pass over the $NK$ design coordinates is conducted with no improvements found. 

\begin{algorithm}[htbp]
\caption{Element-wise CEXCH Pseudocode}
	  \label{alg:cexch}
%\textbf{Algorithm} Compete, steps of 8 pt are used for loop indents
  \begin{algorithmic}[1]
	  \STATE \hspace{0pt}{\textbf{Inputs:} $K:=$ number of experimental factors, $N:=$ number of affordable experimental runs, $f:=$ the optimal design criterion}
		\vspace{0.1in}		
		\STATE \hspace{0pt}{\textbf{//} Randomly instantiate an $N\times K$ design matrix from a uniform distribution}
		\STATE \hspace{0pt}{$\mathbf{X} \leftarrow \{x_{ij} \overset{i.i.d}{\sim} U(-1,1) \text{ for } i = 1, \hdots, N, j = 1, \hdots, K \}$}
		\STATE \hspace{0pt}{\textbf{while }{\{\textit{improvements to $f$ are found} \} \textbf{do}}} 
		\STATE \hspace{8pt}{\textbf{for} $i = 1, 2, \hdots, N$ \textbf{do} }
		\STATE \hspace{16pt}{\textbf{for} $j = 1, 2, \hdots, K$ \textbf{do} }
		\STATE \hspace{24pt}{$\mathbf{X} \leftarrow \argmax_{x_{ij} \in [-1, 1]} f(x_{ij} | \mathbf{X}_{-ij})$ \quad // coordinate proposal is a univariate optimization} \\ \label{optim:kvar:calc}
		\STATE \hspace{16pt}{\textbf{endfor}}
		\STATE \hspace{8pt}{\textbf{endfor}}
		\STATE \hspace{0pt}\textbf{endwhile}
    \vspace{0.1in}	
		\STATE \hspace{0pt}{\textbf{Output:} $\mathbf{X}:=$ a locally optimal design}
  \end{algorithmic}
\end{algorithm}

\subsection{Row-wise EXCH}
This algorithm represents the middle-ground of greediness for all algorithms in this study. \cite{stallrich} provides some discussion of row-exchange algorithms and their application/tractability on optimal design problems. In this algorithm, one only loops through the rows of design $\mathbf{X}$ and exchanges (under some proposal mechanism) a $K$-variate design point $\mathbf{x}_i'$ in an effort to improve the candidate design's criterion score. In Line \ref{optim:univar:calc} of Algorithm \ref{alg:rexch}, we implement the $K$-variate \texttt{BOBYQA} optimization algorithm from the \texttt{NLOpt} package of \texttt{Julia}. The algorithm proceeds by looping through the rows of $\mathbf{X}$ and repeating the $K$-variate optimization until a full pass over the $N$ design points is conducted with no improvements found. 

\begin{algorithm}[htbp]
\caption{Row-wise EXCH Pseudocode}
\label{alg:rexch}
%\textbf{Algorithm} Compete, steps of 8 pt are used for loop indents
  \begin{algorithmic}[1]
	  \STATE \hspace{0pt}{\textbf{Inputs:} $K:=$ number of experimental factors, $N:=$ number of affordable experimental runs, $f:=$ the optimal design criterion}
		\vspace{0.1in}		
		\STATE \hspace{0pt}{\textbf{//} Randomly instantiate an $N\times K$ design matrix from a uniform distribution}
		\STATE \hspace{0pt}{$\mathbf{X} \leftarrow \{x_{ij} \overset{i.i.d}{\sim} U(-1,1) \text{ for } i = 1, \hdots, N, j = 1, \hdots, K\}$}
		\STATE \hspace{0pt}{\textbf{while }{\{\textit{improvements to $f$ are found} \} \textbf{do}}} 
		\STATE \hspace{8pt}{\textbf{for} $i = 1, 2, \hdots, N$ \textbf{do} }
		\STATE \hspace{16pt}{$\mathbf{X} \leftarrow \argmax_{\mathbf{x}'_i \in [-1, 1]^K} f(\mathbf{x}'_i | \mathbf{X}_{-i})$ \quad // design-point proposal is a $K$-variate optimization} \\ \label{optim:univar:calc}
		\STATE \hspace{8pt}{\textbf{endfor}}
		\STATE \hspace{0pt}\textbf{endwhile}
    \vspace{0.1in}	
		\STATE \hspace{0pt}{\textbf{Output:} $\mathbf{X}:=$ a locally optimal design}
  \end{algorithmic}
\end{algorithm}

\section{Results}\label{}
We summarize the efficacy of algorithm performance in two measures: (1) $\hat{\Pr}(\textit{generating a 95\% efficient or better design})$ and (2) $\hat{\Pr}(\textit{generating the best-known design})$ similar to data presented by \cite{walsh1}. Estimates of each probability are computed as the proportion of times in $n_{run}$ we observed the success result in our data. 

\begin{figure}[h!]
  \centering
  \includegraphics[width = 4in]{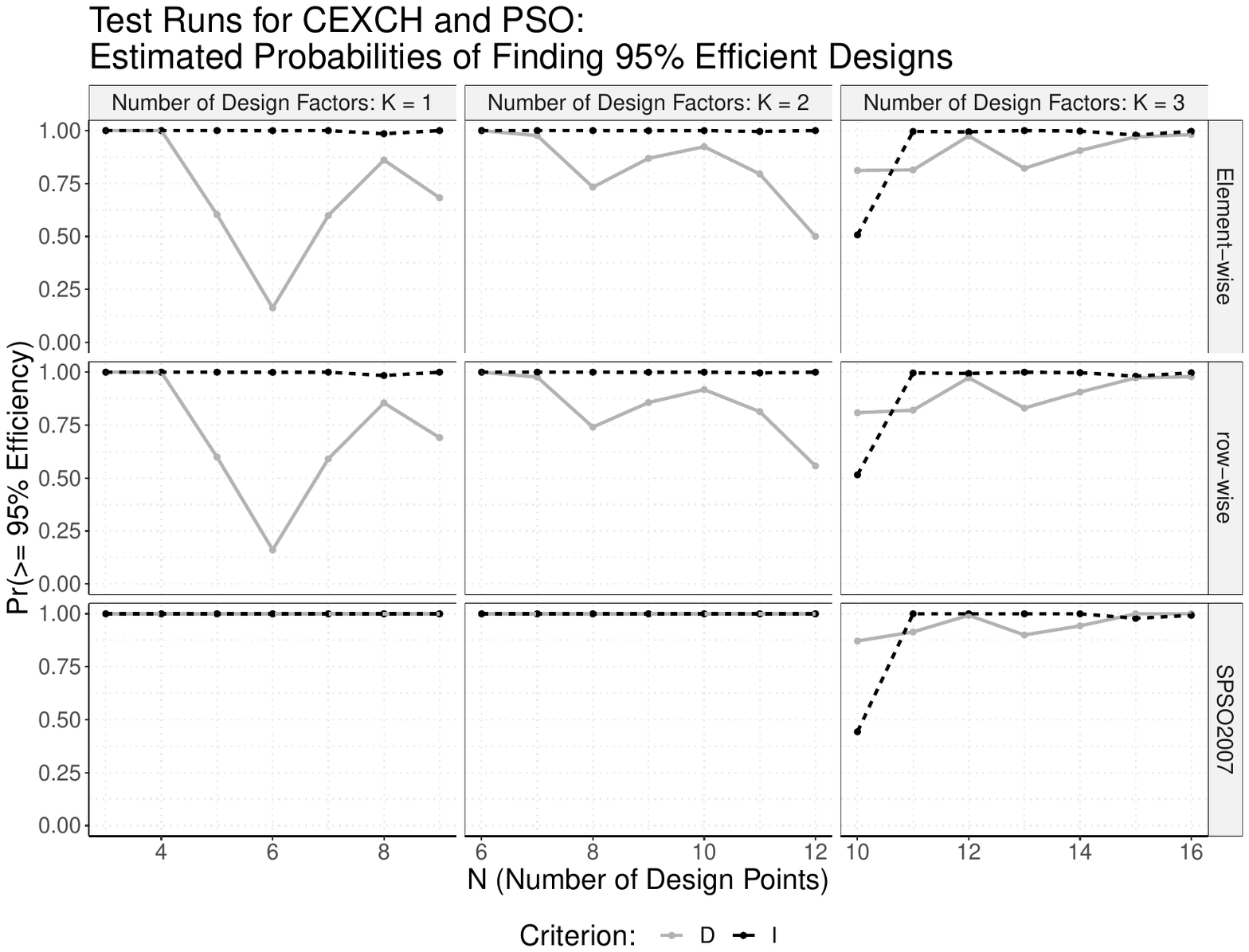}
	  \caption{Data driven estimates that each algorithm generates a 95\% efficient or better design in single run for each scenario.}
		  \label{fig:p95}
\end{figure}

First we briefly summarize Figure \ref{fig:p95} which reports probability estimates for each algorithm generating a 95\% efficient or better design. Rows of the graph panel represent algorithm, columns represent $K$ (\# design factors), and the horizontal axis represents $N$, the number of experimental runs. The black dotted line represents probability estimates for generating $I$ optimal designs. We essentially observe no difference in performance between any of the algorithms for generating $I$-optimal design over all $\{K, N\}$ design scenarios. Switching to $D$-optimal designs and the grey curve, we can SEE PSO demonstrates appreciably larger probability of finding highly efficient designs for the $K = 1, 2$ scenarios, but not for $K = 3$, indicating PSO would need stronger parameters to realize a benefit. Last, we note that element-wise CEXCH (least-greedy) and row-wise (medium greedy) appear to exhibit the same efficacy, a result that seems in-contra to the results of \cite{cexch}.

\begin{figure}[h!]
  \centering
  \includegraphics[width = 4in]{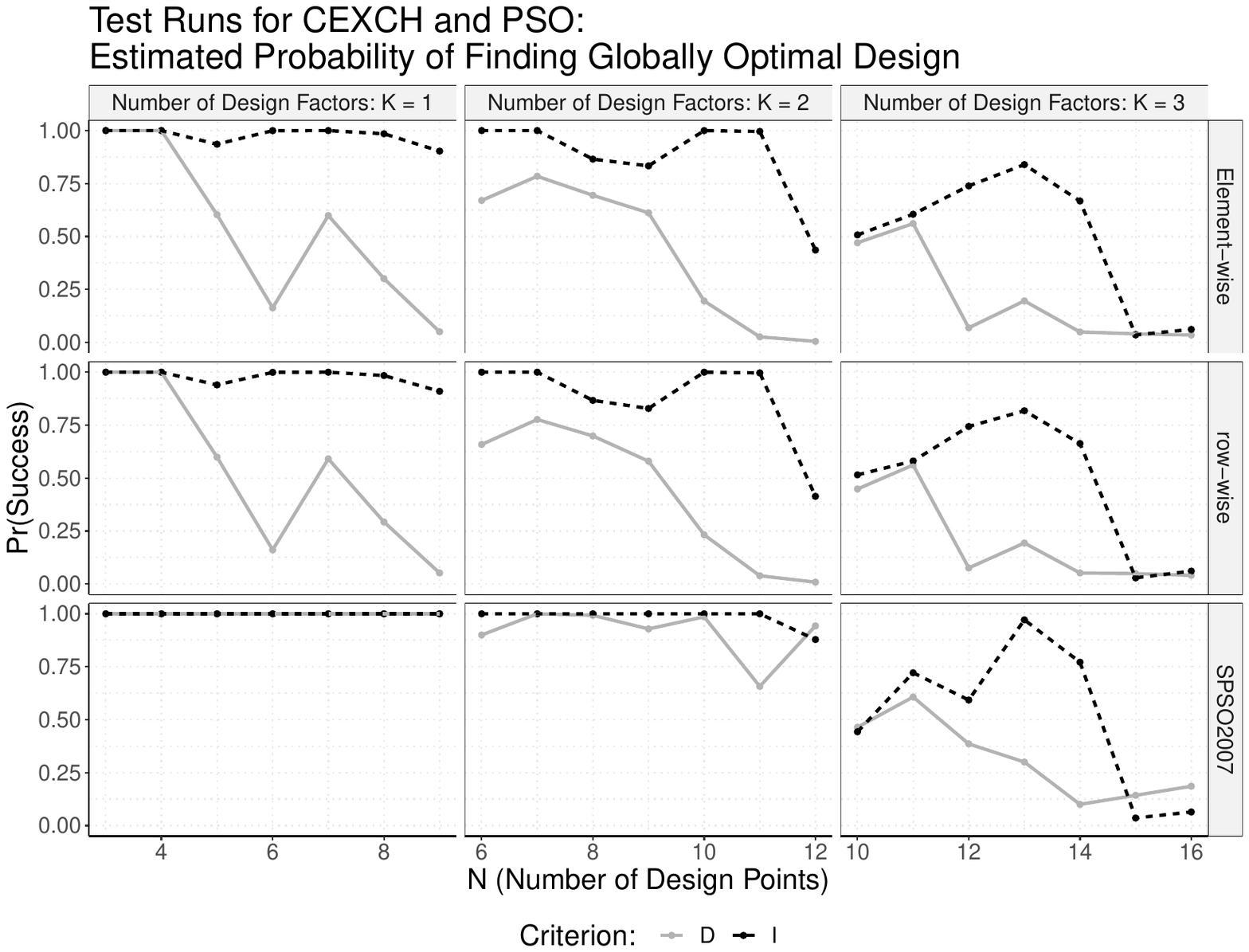}
	  \caption{Data driven estimates that each algorithm generates the globally optimal design in single run for each scenario.}
		  \label{fig:p100}
\end{figure}

Next, we inspect each algorithms efficacy in finding the globally optimal design and present probability of success estimates in Figure \ref{fig:p100}. First, inspecting the black curve, $I$-optimal design searches, for $K= 1$ PSO found the globally optimal design each time, but each CEXCH found it with high probability. PSO appears to have a small edge on the EXCH algorithms for $K$= 2, and for $K$=3, PSO does better for some $N$ but not for others. Last, the results of element-wise and row-wise CEXCH appear to be essentially the same. Next, switching to the grey curves for $D$-optimal design searches, PSO does appreciably better than the EXCH algorithms in finding the globally optimal design for the $K = 1, 2$ scenarios, and only slightly better for the $K = 3$ scenarios. Again, we observe no appreciable difference in performance between the most greedy and less greedy exchange algorithms.

\section{Conclusions and Future Work}\label{}
In this study we apparently found results somewhat in contra to the initial observations and motivations of \cite{cexch}. There may be several explanations. First, \cite{cexch} provided benchmarking data comparing CEXCH to the older PEXCH algorithms which utilize a pre-defined candidate set and apparently only published a time-cost comparison. In addition, \cite{cexch} apparently only benchmarked $D$-optimal design runs, and it seems necessary that such benchmarking be conducted over several optimal design objectives which have different characteristics. Further, when comparing the quality of designs returned by the two algorithms, \cite{cexch} (pp.64 \S3) state \emph{``...Although it is common practice to run a design algorithm from several starting designs, each test case in the current study was run just once.''} To our view it is necessary to study and benchmark algorithm performance on many independent runs of the algorithm. Seemingly in contrast to \cite{cexch}, the most greedy element-wise and less greey row-wise EXCH algorithms appear to have the smae performance when using optimizers as the exchange algorithm. This result should be investigated further. Further, any algorithm appeard to do an equally good job at generating $I$-optimal designs, but the differences expected between PSO and CEXCH were more apparent with the $D$-criterion. This suggests that, perhaps, the $I$-criterion is not so fraught with local optima as the $D$-optimal designs may be. Another result that should be investigated further. 

These results suggests that our community should revisit the greediness dimension in algorithms for generating optimal designs. In future work we intend to expand this study in several dimensions: 1) a more comprehensive model set, including first-order, interaction, linear and non-linear models, 2) larger experiments (i.e. increasing $K$), and 3) a broader set of optimal design criteria including $A$- and $G-$criterion, and 4) inclusion of additional versions of CEXCH and other metaheursitics that have been used for generating optimal DoEs (sp. GA and SA). 


\newpage

\bibliographystyle{alpha}

\bibliography{mybib}

\newcommand{\etalchar}[1]{$^{#1}$}
\begin{thebibliography}{WCHW15}

\bibitem[ADT07]{adt}
A.C. {Atkinson}, A.N. {Donev}, and R.D. {Tobias}.
\newblock {\em Optimal Experimental Designs, with SAS}.
\newblock Oxford Statistical Science Series, 2007.

\bibitem[BEKS17]{julia}
Jeff Bezanson, Alan Edelman, Stefan Karpinski, and Viral~B Shah.
\newblock Julia: A fresh approach to numerical computing.
\newblock {\em SIAM review}, 59(1):65--98, 2017.

\bibitem[Bor03]{jobo1}
{John J.} Borkowski.
\newblock Using a genetic algorithm to generate small exact response surface
  designs.
\newblock {\em Journal of Probability and Statistical Science}, 1(1), Feb.
  2003.

\bibitem[CCW{\etalchar{+}}15]{chen3}
Ray-Bing Chen, Shin-Perng Chang, Weichung Wang, Heng-Chih Tung, and Weng~Kee
  Wong.
\newblock Minimax optimal designs via particle swarm optimization methods.
\newblock {\em Statistics and Computing}, 25(5):975--988, Sep 2015.

\bibitem[CCW22]{psorev}
Ping-Yang Chen, Ray-Bing Chen, and Weng~Kee Wong.
\newblock Particle swarm optimization for searching efficient experimental
  designs: A review.
\newblock {\em WIREs Computational Statistics}, 14(5):e1578, 2022.

\bibitem[CHHW13]{chen2}
Ray-Bing Chen, Dai-Ni Hsieh, Ying Hung, and Weichung Wang.
\newblock Optimizing latin hypercube designs by particle swarm.
\newblock {\em Statistics and Computing}, 23(5):663--676, Sep 2013.

\bibitem[CHHW14]{chen1}
R-B. {Chen}, Y-W. {Hsu}, Y.~{Hung}, and W.~{Wang}.
\newblock Discrete particle swarm optimization for constructing uniform design
  on irregular regions.
\newblock {\em Computational Statistics and Data Analysis}, 72:282--297, 2014.

\bibitem[CLHW19]{chen4}
Ray-Bing Chen, Chi-Hao Li, Ying Hung, and Weichung Wang.
\newblock Optimal noncollapsing space-filling designs for irregular
  experimental regions.
\newblock {\em Journal of Computational and Graphical Statistics},
  28(1):74--91, 2019.

\bibitem[GJ11]{odt}
P.~{Goos} and B.~{Jones}.
\newblock {\em Optimal Design of Experiments: A Case study approach.}
\newblock John Wiley and Sons, Ltd., 2011.

\bibitem[HN18]{gIopt}
Lucia~N. Hernandez and Christopher~J. Nachtsheim.
\newblock Fast computation of exact {$G$}-optimal designs via
  {$I_{\lambda}$}-optimality.
\newblock {\em Technometrics}, 60(3):297--305, 2018.

\bibitem[JN83]{cexch2}
Mark~E. Johnson and Christopher~J. Nachtsheim.
\newblock Some guidelines for constructing exact {$D$}-optimal designs on
  convex design spaces.
\newblock {\em Technometrics}, 25(3):271--277, 1983.

\bibitem[Joh]{nlopt}
S.G. Johnson.
\newblock The nlopt nonlinear-optimization package.

\bibitem[LMW19]{lukemire}
Joshua Lukemire, Abhyuday Mandal, and Weng~Kee Wong.
\newblock d-qpso: A quantum-behaved particle swarm technique for finding
  {$D$}-optimal designs with discrete and continuous factors and a binary
  response.
\newblock {\em Technometrics}, 61(1):77--87, 2019.

\bibitem[MJ18]{joseph}
Simon Mak and V.~Roshan Joseph.
\newblock Minimax and minimax projection designs using clustering.
\newblock {\em Journal of Computational and Graphical Statistics},
  27(1):166--178, 2018.

\bibitem[MMA16]{rsm}
R.~{Myers}, D.~{Montgomery}, and C.~{Anderson-Cook}.
\newblock {\em Response surface methodology: process and product optimization
  using Designed Experiments. 4th Edition.}
\newblock John Wiley and Sons, Ltd., 2016.

\bibitem[MN95]{cexch}
Ruth~K. Meyer and Christopher~J. Nachtsheim.
\newblock The coordinate-exchange algorithm for constructing exact optimal
  experimental designs.
\newblock {\em Technometrics}, 37(1):60--69, 1995.

\bibitem[RJBM10]{rodman}
Myrta Rodríguez, Bradley Jones, Connie~M. Borror, and Douglas~C. Montgomery.
\newblock Generating and assessing exact {$G$}-optimal designs.
\newblock {\em Journal of Quality Technology}, 42(1):3--20, 2010.

\bibitem[SAMJ23]{stallrich}
Jonathan Stallrich, Katherine Allen-Moyer, and Bradley Jones.
\newblock D- and a-optimal screening designs.
\newblock {\em Technometrics}, 0(0):1--10, 2023.

\bibitem[SP16]{saleh}
Moein Saleh and Rong Pan.
\newblock A clustering-based coordinate exchange algorithm for generating
  g-optimal experimental designs.
\newblock {\em Journal of Statistical Computation and Simulation},
  86(8):1582--1604, 2016.

\bibitem[Wal21]{walsh}
S.~J. Walsh.
\newblock {\em Development and applications of particle swarm optimization for
  constructing optimal experimental designs.}
\newblock Ph.D. Thesis. Montana State University., 2021.

\bibitem[WB22a]{walsh1}
Stephen~J. Walsh and John~J. Borkowski.
\newblock Generating exact optimal designs via particle swarm optimization:
  Assessing efficacy and efficiency via case study.
\newblock {\em Quality Engineering}, 35(2):304--323, 2022.

\bibitem[WB22b]{walsh2}
Stephen~J. Walsh and John~J. Borkowski.
\newblock Improved g-optimal designs for small exact response surface
  scenarios: Fast and efficient generation via particle swarm optimization.
\newblock {\em Mathematics}, 10(22), 2022.

\bibitem[WCHW15]{wong}
Weng~Kee Wong, Ray-Bing Chen, Chien-Chih Huang, and Weichung Wang.
\newblock A modified particle swarm optimization technique for finding optimal
  designs for mixture models.
\newblock {\em PLOS ONE}, 10(6):1--23, 06 2015.

\bibitem[WLAC23]{walsh3}
Stephen~J. Walsh, Lu~Lu, and Christine~M. Anderson-Cook.
\newblock I-optimal or g-optimal: Do we have to choose?
\newblock {\em Quality Engineering}, 0(0):1--22, 2023.

\end{thebibliography}



\end{document}